%% file: scalar.tex
\global\def\draftcontrol{0}
   \def\versionno{ Scalar Collapse in AdS  }
\input{alex.tex}
\documentclass[12pt,letterpaper]{article}

\usepackage{amsmath,amssymb,array,calc,epsfig}
\usepackage{graphicx}
\usepackage{psfrag,verbatim,bm}
\usepackage{hyperref}
\usepackage{amssymb,graphicx}
\usepackage[usenames]{color}

\begin{document}
\input{alex2.tex}


\title{\bf Scalar Collapse in AdS}
\pubnum
{UWO-TH-12/10
}

\date{October 1, 2012}

\author{
Alex Buchel,$^{1,2}$ Luis Lehner$^{1,3}$ and Steven L. Liebling$^{1,4}$\\[0.4cm]
\it $^1$\,Perimeter Institute for Theoretical Physics\\
\it Waterloo, Ontario N2J 2W9, Canada\\[0.2cm]
\it $^2$\,Department of Applied Mathematics\\
\it University of Western Ontario\\
\it London, Ontario N6A 5B7, Canada\\[0.2cm]
\it $^3$\,Department of Physics, University of  Guelph\\
\it Guelph, Ontario N1G 2W1, Canada\\[0.2cm]
\it $^4$\,Department of Physics, Long Island University\\
\it Brookville, NY 11548, U.S.A.\\[0.2cm]
}

\Abstract{
Recently, studies of the gravitational collapse of a scalar field within
spherically symmetric AdS spacetimes was presented in ~\cite{Bizon:2011gg,Jalmuzna:2011qw} which showed
an instability of pure AdS to black hole formation. In particular,
the work showed that arbitrarily small initial configurations of
scalar field evolved through some number of reflections off the AdS
boundary until a black hole forms. We consider this same system,
extended to include a complex scalar field, and reproduce this 
phenomena. We present tests of our numerical code that demonstrate
convergence and consistency. We study the properties of the evolution as the scalar pulse
becomes more compact examining the asymptotic behavior of the scalar
field, an observable in the corresponding boundary CFT.
We demonstrate that such BH formation occurs
even when one places a reflecting boundary at finite radius indicating
that the sharpening is a property of gravity in a bounded domain, not of
AdS itself. We examine how the initial energy is transferred to higher frequencies --which
leads to black hole formation-- and uncover interesting features of this transfer.
}

\makepapertitle

\body

\version\versionno

\section{Introduction}

Two recent papers~\cite{Bizon:2011gg,Jalmuzna:2011qw}
 study the gravitational collapse of a real scalar field in spherically
symmetric Anti-de Sitter~(AdS) spacetimes finding a remarkable instability
to black hole~(BH) formation. That is, given any non-vanishing
initial scalar field, they provide convincing numerical evidence, further
supported by analytic arguments, 
 that
the resulting spacetime forms a BH within a time that scales inversely
with the square of the initial amplitude of the scalar perturbation to AdS\footnote{It is interesting 
to note that Pretorius \& Choptuik foresaw this
possibility in their study in $2+1$ AAdS~\cite{Pretorius:2000yu}.}.
The work discussing this very interesting result, however,
presents relatively few details and results from related work~\cite{Garfinkle:2011hm,Garfinkle:2011tc} 
are in tension with it. We have therefore reproduced aspects of~\cite{Bizon:2011gg,Jalmuzna:2011qw}
and provide new insights in this paper.

Such a result is, at first sight, surprising from a few different perspectives. 
Since  Minkowski and de Sitter spacetimes are stable, it might be 
expected that AdS is stable as well~\cite{Dias:2011ss}. Within the
context of black hole critical phenomena~\cite{Choptuik:1992jv,Gundlach:2007gc}, 
the result is similarly surprising. In studies of critical behavior, one generally
evolves a series of pulses with different amplitudes
as the initial pulses implode through their centers, searching
for the threshold of black hole formation. However, this instability
suggests that a black hole is {\em always} formed and that this black hole
threshold is only a threshold for {\em immediate} collapse.

In contrast, one can similarly view this instability from the perspective
of the corresponding boundary conformal field theory~(CFT). AdS-CFT
asserts that BH solutions in the bulk correspond to thermal states of the
CFT. In this light, the fact that this instability would always form a black hole 
would simply imply that states thermalize, which is not surprising at all. 
Furthermore, evidence for the instability has also been given in~\cite{Anderson:2006ax,dafermos}
supporting this behavior based on ergodic arguments or linearized perturbation analysis as well
as a theorem guaranteeing the instability under certain conditions~\cite{Ishibashi:2012xk}.
On the other hand, 
recent works indicate that some scenarios never form a black hole, thus thermalization within
the gauge/gravity context need not be a forgone conclusion~\cite{Dias:2012fu,Buchelboson}.

Furthermore, a pure state can never thermalize during a unitary evolution, and so the BH formation 
conjectures of~\cite{Bizon:2011gg,Jalmuzna:2011qw}, as interpreted within the framework of AdS/CFT 
correspondence, indicate that it might be impossible to set-up a pure initial state in 
strict limit large-$N$ gauge theories. Likewise, if the initial configuration carries a 
conserved global charge, one expects (at most) thermalization in a given charge sector.
A global charge in the  boundary CFT dynamics is realized through a local gauge symmetry 
of the dual gravitation bulk dynamics. We postpone the  study of thermalization    
of charged configurations in AdS to future work, and discuss here the effect of
global symmetries in the bulk on the thermalization. To this end we extend the work 
of~\cite{Bizon:2011gg,Jalmuzna:2011qw} to a complex scalar field collapse in AdS. 

This work is organized as follows;
formulation of the problem is given in Section~\ref{sec:formulation}.
We present a number of details of our numerical work in
Section~\ref{sec:numerics} and provide tests that our code
produces convergent and consistent solutions through a number of
reflections off the boundary. We consider the results of our simulations
in Section~\ref{sec:results}, extending to a complex scalar field. In Section~\ref{sec:cft} 
we interpret the results of our simulations in a context of a dual boundary 
conformal gauge theory dynamics. We conclude in Section~\ref{sec:conclude}.

\section{Formulation}
\label{sec:formulation}
Consider a $(d+1)$-dimensional gravitational system, dual to a strongly coupled CFT$_d$
on $(d-1)$-dimensional sphere $S^{d-1}$. 
We focus on $SO(d)$-invariant states of a conformal field theory with nonzero expectation values 
for a pair of exactly marginal operators $\calo_d^{(i)}$, $i=1,2$.     
We are interested in a unitary evolution of such states  as determined by the following dual effective 
gravitational action
\begin{equation}
S_{d+1}=\frac{1}{16\pi G_{d+1}}\int_{\calm_{d+1}}d^{d+1}\xi \sqrt{-g} \left(R_{d+1}+\frac{d(d-1)}{\ell^2}
-2 \del_\mu \phi\del^\mu \phi^*\right)\,,
\eqlabel{acd}
\end{equation}
where $\phi\equiv \phi_1+i\ \phi_2$ is a massless complex scalar field, dual to a pair of exactly marginal
operators $ \calo_d^{(1)}+i\ \calo_d^{(2)}\equiv \calo_d$. 
Further,
\begin{equation}
\calm_{d+1}=\del\calm_{d+1}\times \cali,\qquad \del\calm_{d+1}=R_t\times S^{d-1}\,,\qquad
\cali=\{x\in[0,\frac\pi 2]\}\,.
\eqlabel{manifold}
\end{equation}
Notice that the effective action \eqref{acd} is invariant under a global phase rotation 
\begin{equation}
\phi\to \phi\ e^{-i\a}\,,
\eqlabel{u1symm}
\end{equation}
corresponding to a phase rotation of expectation values 
\begin{equation}
\langle\calo_d\rangle\to \langle\calo_d\rangle\ e^{-i\a}.
\eqlabel{u1symmb}
\end{equation}
Associated with \eqref{u1symm}
there is a conserved current 
\begin{equation}
J^\mu =i g^{\mu\nu } \left(\phi \del_\nu\phi^*-\phi^* \del_\nu\phi\right)\,,
\eqlabel{current}
\end{equation}
and a corresponding conserved charge 
\begin{equation}
Q=\int_{S^{d-1}} dS^{d-1}\int_0^{\pi/2} dx\ \sqrt{-g} J^t .
\eqlabel{q}
\end{equation}
The other conserved quantity of a CFT during the evolution is the total mass~$M$. We can rephrase the 
work~\cite{Bizon:2011gg,Jalmuzna:2011qw} as a statement that nonequilibrium states $\{M,Q=0\}$ of a 
strongly coupled CFT with a gravitational dual evolve into  thermal states. Here, 
in addition to reproducing these claims (and reinterpreting the gravitational bulk dynamics 
from the boundary CFT perspective), we are interested in a broader question: 
{\em what is the role of a non-zero charge $Q$ in the process of thermal equilibration?}
Such a question is particularly interesting as it is well-known that a complex
scalar field in AdS supports perturbatively-stable boson star solutions~\cite{Astefanesei:2003qy}.
    
To proceed, we adopt a notation similar to that of~\cite{Jalmuzna:2011qw} choosing
the same form of the $(d+1)$-dimensional metric describing an asymptotically
AdS spacetime
\begin{equation}
ds^2 = \frac{\ell^2}{\cos^2 x} \left(
                                     -Ae^{-2\delta} dt^2
                                     +\frac{dx^2}{A}
                                     +\sin^2x \, d\Omega^2_{d-1}
                                        \right) \, ,
\eqlabel{eq:metric}
\end{equation}
where $\ell$ is the scale-size of the AdS spacetime,
$d\Omega^2_{d-1}$ is the metric of $S^{d-1}$, and $A(x,t)$ and $\delta(x,t)$ 
are scalar functions describing the metric. The scale $\ell$ drops
out of the resulting equations and we set it to unity without loss of
generality. The spatial coordinate $x$ runs over $0\rightarrow \pi/2$ such that
spheres through the point $x$ have a radius $r \equiv \tan x$.
To describe  the real and imaginary components of a scalar field, we introduce the two quantities
$\Phi_i \equiv \left(\partial/\partial x\right) \left( \phi_i \right)$
and
$\Pi_i \equiv A^{-1} e^\delta \left(\partial/\partial t\right) \left( \phi_i \right)$.

Besides the extension from a real scalar
to a complex scalar field, this is precisely as described in~\cite{Jalmuzna:2011qw} and we derive 
the same equations of motion. However, working with these
variables, achieving a stable numerical scheme required a few numerical
tweaks at the outer boundary to provide  convergent
evolutions even through reflections (or bounces). A more
elegant solution, and one which requires no such tricks, involves rescaling
the matter functions in keeping with their appropriate boundary fall-off.
In particular, we use rescaled quantities according to
\begin{eqnarray}
\hat \phi_i & \equiv  & \frac{\phi_i}{\cos^{d-1}x} \, , \\
\hat \Pi_i  & \equiv  & \frac{e^\delta}{A}\frac{\partial_t \phi_i}{\cos^{d-1}x} 
                  = \frac{\Pi_i}{\cos^{d-1}x} \, ,\\ 
\hat \Phi_i & \equiv  & \frac{\partial_x\phi_i}{\cos^{d-2}x}
                  = \frac{\Phi_i}{\cos^{d-2}x} \, .
\end{eqnarray}

In terms of these rescaled quantities (we drop the caret from here forward), we have
the following evolution equations resulting from the Klein-Gordon equation
\begin{equation}
\begin{split}
\dot \phi_i  = & A e^{-\delta} \Pi_i\,, \\
\dot \Phi_i  = & \frac{1}{\cos^{d-2}x} \left( \cos^{d-1}x A e^{-\delta} \Pi_i \right)_{,x}\,,\\
\dot \Pi_i   = & \frac{1}{\sin^{d-1}x} \left( \frac{\sin^{d-1}x}{\cos x}
                                               A e^{-\delta} \Phi_i
                                               \right)_{,x}\,.
\end{split}
\eqlabel{kg}
\end{equation}
The system also includes  two spatial equations to be integrated
\begin{equation}
\begin{split}
A_{,x}       = &   \frac{d-2 + 2 \sin^2x}{\sin x \cos x} \left(1-A\right)
                  - \sin x \cos^{2d-1}x A\left( \frac{\Phi_i^2}{\cos^2x} + \Pi_i^2 \right)\,,\\
\delta_{,x}  = &  - \sin x \cos^{2d-1}x \left( \frac{\Phi_i^2}{\cos^2x}+ \Pi_i^2 \right)\,,
\end{split}
\eqlabel{constx}
\end{equation}
together with one constraint equation
\begin{equation}
A_{,t} + 2 \sin x \cos^{2d-2} A^2 e^{-\delta} \left( \Phi_i \Pi_i \right)=0\,,
\eqlabel{consteq}
\end{equation}
where a sum over $i= \{1,2\}$ is implied.\\

At the origin, these quantities behave independently of $d$ as
\begin{equation}
\begin{split}
\phi_i(t,x)  = & \phi_0^{(i)}(t) + {\cal O}(x^2)\,,  \\
A(t,x)       = &  1 + {\cal O}(x^2)\,,  \\
\delta(t,x)  = &  \delta_0(t) + {\cal O}(x^2)\,.
\end{split}
\eqlabel{ir}
\end{equation}
At the outer boundary $x=\pi/2$ we introduce $\rho \equiv \pi/2-x$ so that we have
\begin{equation}
\begin{split}
\phi_i(t,\rho)  = & \phi_d^{(i)}(t)\rho + {\cal O}(\rho^3)\,,  \\
A(t,\rho)       = &  1 - M \frac{\sin^d\rho}{\cos^{d-2}\r} 
+  {\cal O}(\rho^{2 d})\,,  \\
\delta(t,\rho)  = &  0 + {\cal O}(\rho^{2d})\,.
\end{split}
\eqlabel{uv}
\end{equation}
We note here a difference between our work and~\cite{Jalmuzna:2011qw}. We choose
conditions on $\delta$ such that it becomes zero on the boundary and hence the
coordinate time with which we examine the boundary describes proper time there as well.
The gauge freedom in this system allows for such a rescaling without affecting
the dynamics observed.

The asymptotic behavior \eqref{uv} determines the boundary CFT observables: 
the expectation values of the stress-energy tensor $T_{kl}$, and the
operators $\calo_{d}^{(i)}$ 
\begin{equation}
\begin{split}
&8\pi G_{d+1}\langle T_{tt}\rangle =\cale_d\,,\qquad 
\langle T_{\a\b}\rangle= \frac{g_{\a\b}}{d-1}\ \langle T_{tt}\rangle\,,\\
&16\pi G_{d+1}\langle \calo_{d}^{(i)} \rangle=4 d\ \phi_d^{(i)}(t)\,,
\end{split}
\eqlabel{vevs}
\end{equation}
where $\cale_d$, up to an additive constant\footnote{This constant
determines the Casimir energy of the CFT$_d$.} for even $d$, is proportional 
to $M$; $g_{\a\b}$ is a metric on a round $S^{d-1}$. Explicitly, for 
$d=3,4$ we have
\begin{equation}
\cale_d=\begin{cases}
M\,,& \qquad d=3\,, \\
\frac 38 (1+4 M)\,,& \qquad d=4\,.
\end{cases}
\eqlabel{ed}
\end{equation}
Additionally note that the conserved $U(1)$ charge is given by 
\begin{equation}
Q=4\pi^2\ \int_0^{\pi/2} dx\ 
\sin^{d-1}x\ \cos^{d-1}x\ \left(\Pi_1(0,x) \phi_2(0,x)-\Pi_2(0,x)\phi_1(0,x)\right)  .
\eqlabel{qinit}
\end{equation} 
Note that since $\del_t Q=0$, the integral in \eqref{qinit} can be evaluated at 
$t=0$.

The constraint \eqref{consteq} implies that $M$ in \eqref{uv} 
is  time-independent, ensuring            energy
conservation
\begin{equation}
\del_t\ \cale_d=0\,.
\eqlabel{emc} 
\end{equation}

It is convenient to introduce the mass aspect function $\calm(t,x)$
as 
\begin{equation}
A(t,x)=1 - \calm(t,x) \frac{\cos^d x}{\sin^{d-2} x}\, .
\eqlabel{massfunction}
\end{equation}
Following \eqref{constx} we find 
\begin{equation}
\calm(t,x)=\int_0^x dz\ \tan^{d-1} z\ \cos^{2(d-1)}z\  A(t,z) 
\left[\frac{\Phi_i^2(t,z)}{\cos^2 z}+\Pi_i^2(t,z)\right]\,.
\eqlabel{mf1}
\end{equation}
Comparing \eqref{mf1} and \eqref{uv} we see that
\begin{equation}
M=\calm(t,x)\bigg|_{x=\frac\pi2}\,.
\eqlabel{mcalm}
\end{equation}

\begin{figure}[t]
\begin{center}
\psfrag{m}{{$\ln M$}}
\psfrag{e}{{$\ln \epsilon$}}
\includegraphics[width=3in]{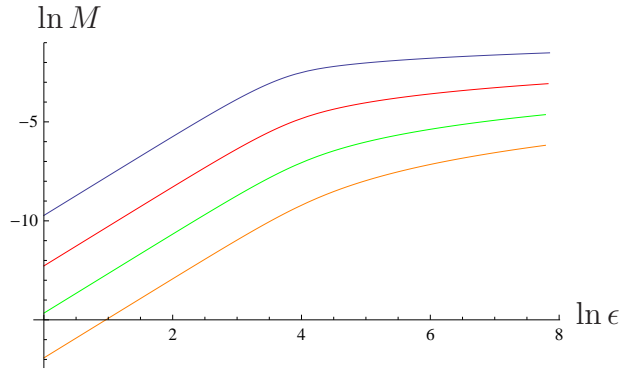}
\end{center}
  \caption{ Mass $M$ of initial configurations \eqref{phiQz} as a function of 
$\e$ for $d={3,4,5,6}$ (blue/red/green/orange curves). Note that $M\propto \e^2$ for 
 $\ln\e\lesssim 4$. While the coordinate extent of the scalar profile is the same
in all $d$, higher dimensional $AdS_{d+1}$ more efficiently ``localizes'' it, resulting 
in smaller $M_d$ for a fixed $\e$.  
}\label{afig1}
\end{figure}

\begin{figure}[t]
\begin{center}
\psfrag{m}{{$\ln M$}}
\psfrag{e}{{$\ln\epsilon$}}
\psfrag{ms}{{$5^{d-3}M$}}
\psfrag{mq}{{$\ln\frac{M}{Q}$}}
  \includegraphics[width=2.4in]{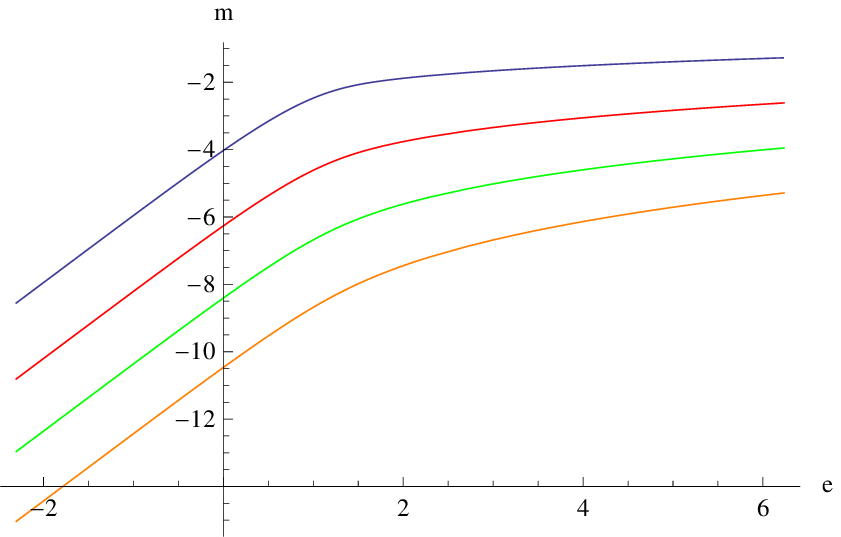}
  ~~~
  \includegraphics[width=2.4in]{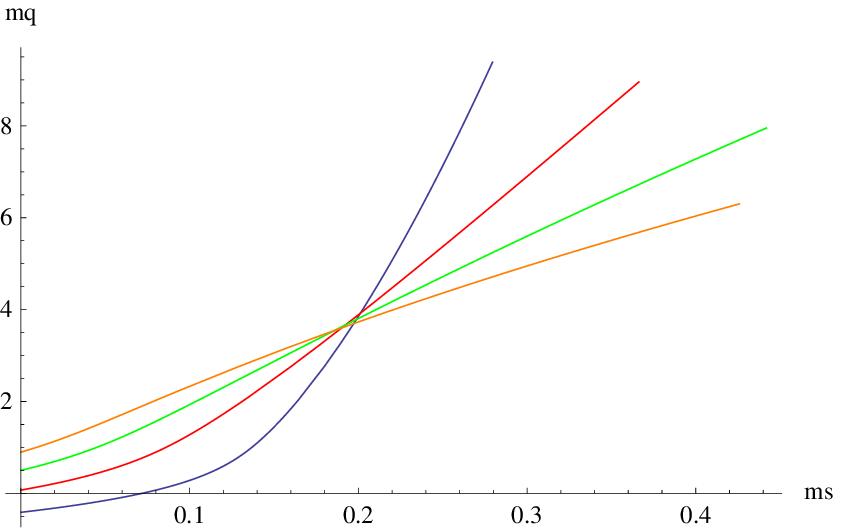}
\end{center}
  \caption{ Mass $M$ of initial configurations \eqref{phiQnz} as a function of 
$\e$ for $d=3\cdots6$ (left panel, blue$\cdots$orange) and corresponding ratios
$\frac{M}{Q}$ (right panel). Here, $M\propto \e^2$ scaling of mass extends only 
up to $\ln\e \sim 1$. Note that $\frac{M}{Q}$ approaches a constant as $M\to 0$,
while $\ln\frac{M}{Q}\propto M$ for large $M$, effectively making the charge negligible.
The rescaling of the horizontal axis of the right panel by $5^{d-3}$ is made for readability
of the graph.
}\label{afig2}
\end{figure}

Both $\cale_d$ and the charge $Q$ are determined from the initial data.
Motivated by~\cite{Bizon:2011gg}, in this paper we consider the 
following set of initial conditions:
\nxt $Q=0$ case
\begin{equation}
\Phi_i(0,x)=0\,,\qquad \Pi_i(0,x)=\frac{2\epsilon}{\pi}
e^{-\frac{4\tan^2x}{\pi^2\sigma^2}}\ \cos^{1-d}x\  \delta_i^1\,,
\eqlabel{phiQz}
\end{equation}
where here $\delta^j_k$ represents the Kronecker delta.
Numerically solving the spatial ODEs of \eqref{massfunction} and \eqref{mf1},
we compute $\calm(0,\pi/2)=M$. For $d=3\cdots6$ and $\sigma=\frac{1}{16}$ 
the values $M(\epsilon)$ are presented in Fig.~\ref{afig1}. 
\nxt $Q\ne 0$ case
\begin{equation}
\begin{split}
&\Phi_i(0,x)=\del_x\left[\frac{2\epsilon}{\pi}
e^{-\frac{4\tan^2x}{\pi^2\sigma^2}}\ \cos^{1-d}x\ \delta_i^1\right]\,,\\
&\Pi_i(0,x)=\w_d\ \frac{2\epsilon}{\pi}
e^{-\frac{4\tan^2x}{\pi^2\sigma^2}}\ \cos^{1-d}x\ \delta_i^2\,,\qquad 
\w_d=d\,,
\end{split}
\eqlabel{phiQnz}
\end{equation}
where $\w_d$ is the lowest frequency of the linearized scalar perturbations of $AdS_{d+1}$.   
Numerically solving the spatial ODEs of \eqref{massfunction} and \eqref{mf1},
we compute $\calm(0,\pi/2)=M$. For $d=3\cdots6$ and $\sigma=\frac{1}{16}$ ,
the values $M(\epsilon)$ and $\frac{M}{Q}$ are presented in Fig.~\ref{afig2}.

Note that for both classes of initial conditions, \eqref{phiQz} and \eqref{phiQnz},
the values $\phi_d^{(i)}(0)$, and correspondingly the expectation values of the boundary 
CFT operators $\calo_d^{(i)}$ at $t=0$, are zero.

\section{Numerics}
\label{sec:numerics}

\subsection{Implementation}
\label{sec:implementation}

To solve these equations, we employ finite-difference approximations within a method-of-lines~(MOL)
approach that employs a third-order accurate Runge-Kutte~(RK3) time evolution scheme.
Such an approach computes the spatial derivatives (with a high-order accurate, finite-difference
stencil) and then considers the evolution equations as ordinary-differential equations~(ODEs) in
the time coordinate.

As is standard in studies of critical behavior, we use adaptive mesh refinement~(AMR) to add
resolution where needed and remove it when not needed. In particular, we use Choptuik's {\tt ad}
infrastructure~\cite{Choptuik:1992jv,frontiers}  adapted for our uses here via the changes:
\begin{itemize}
\item{modified the loop for convergence within the iterative Crank-Nicholson scheme to a three-step RK3 update}
\item{replaced second-order accurate spatial derivatives with high-order accuracy spatial derivatives}
\item{modified the AMR-boundary treatment from interpolation for a single point in time to interpolation in
 time of width two points in the parent, with spatial interpolation on the fine level between these two points. This approach follows in spirit the tapered approach of~\cite{Lehner:2005vc}, although it
strictly lacks the higher-order
convergence there.}
\end{itemize}

The numerical grid has two boundaries, the origin and the AdS boundary at $\rho \equiv \pi/2 -x = 0$. To enforce the boundary conditions~(\ref{ir}) at the
origin, we set $\Phi_i(t,0)=0$ and $A(t,0)=1$ and obtain values for $\phi_i(t,0)$ and $\Pi_i(t,0)$ from quadratic fits. Because $\delta(t,x)$ is computed
by numerical integration from the outer boundary, no boundary
condition is needed at the origin. At the $\rho=0$ boundary, we
enforce the conditions~(\ref{uv}) by setting $\phi_i(t,\pi/2)$, $\Pi_i(t,\pi/2)$, and $\Phi_i(t,\pi/2)$ all to zero there. We set the metric functions
$A(t,\pi/2)=1$ and $\delta(t,\pi/2)=0$, the latter chosen to ensure that
coordinate time $t$ measures proper time on the boundary.

\subsection{Tests}
\label{sec:tests}

\begin{figure}
\centerline{\includegraphics[width=4in]{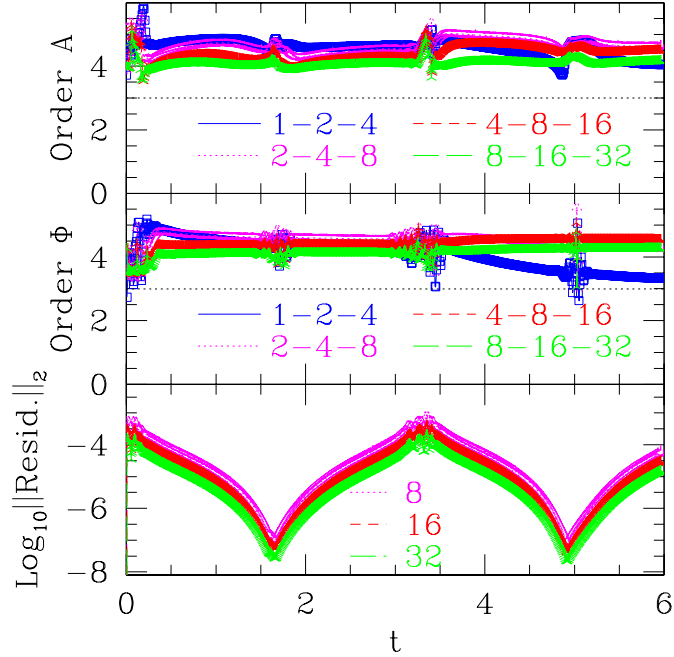}}
\caption{Convergence test for two bounces with $d=3$. The initial data
is of the form~\eqref{phiQz} with $\epsilon=20.01$ and $\sigma=1/16$.
({\bf top}) The
order of convergence obtained from comparisons of the metric function $A(x,t)$
at successively doubled resolutions. The convergence order is computed from three different resolutions, a base resolution and runs with half and one-quarter
the base grid solution. Thus, run ``32'' has a grid spacing $2^{-5}$ that of the run labeled ``1.''({\bf middle}) Order of convergence
obtained for $\Phi_1(x,t)$.  Both these results indicate convergence is better
than third order convergent. ({\bf bottom})~The logarithm of the L2-norm of
the momentum constraint residual for just the three best resolutions.
That it decreases with
increasing resolution suggests the results are converging to a proper solution
of Einstein's equations. Note that this  residual is computed only at
first order of accuracy because it involves a time derivative and therefore
one should {\em not} estimate the order of convergence from it.
}
\label{fig:convergence}
\end{figure}

We carry-out  a number of tests of our implementation. We find that the solutions
obtained generally
converge at least as fast as third-order in the grid spacing. We evaluate
the residual of the momentum constraint and confirm that it approaches zero
as the resolution is increased. We compute both the total mass $\calm(t,\pi/2)$ using~\eqref{mf1} and charge using~\eqref{qinit},
and ensure that their degree of conservation in time improves with resolution.
We also completed two standard black hole threshold searches and find
the expected Choptuik behavior~\cite{Choptuik:1992jv}.
Fig.~\ref{fig:convergence} presents some results from one such test.
For the AMR evolutions mentioned here, we ensure that increasing both the coarse level resolution and decreasing the 
threshold for refinement produce essentially unchanged collapse times.

\section{Results}
\label{sec:results}
\begin{figure}
\centerline{\includegraphics[width=4in]{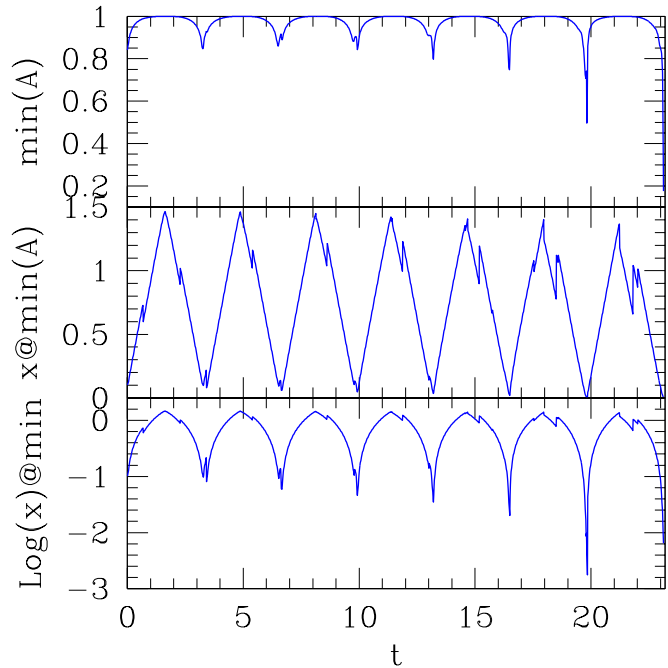}}
\caption{Demonstration of the increasingly compact gravitational potential
with successive bounces of initial data of form~\eqref{phiQz} with $\epsilon=20$ and 
$\sigma=1/16$. Six bounces are shown before a black hole forms.
({\bf top}) The minimum of $A(x,t)$ as a function of time. At the final time,
the minimum approaches zero, signaling black hole formation.
({\bf middle}) The $x$-coordinate where $A(x,t)$ achieves its minimum. In these
coordinates, the speed of light is unity, and the line segments roughly indicate
the motion of the scalar pulse back and forth across the grid. That compaction
increases with each bounce is quite hard to observe on the linear scale, and
so the logarithm of this data is shown at {\bf bottom}. The evolution terminates
at the last time shown as the solution approaches black hole formation.
}
\label{fig:bounces}
\end{figure}

We find the same behavior as found in~\cite{Bizon:2011gg,Jalmuzna:2011qw}.
In particular, given any initial, non-vanishing configuration of scalar field,
an evolution with sufficient resolution and time produces a black hole.
We observe that as the pulse travels back and forth from the boundary,
one sees increasing compactness in the scalar profile. Examining the
metric function $A(x,t)$, one sees that on return of the pulse to the centre,
the metric function approaches 0 to an increasing degree, and it is
precisely the approach of $A(x,t)$ to zero that indicates the formation of a black hole.
In other words,
the increasingly compact scalar profile produces an increasingly deeper
gravitational potential well. The location of this minimum also decreases
with each bounce. We plot these two quantities in Fig.~\ref{fig:bounces}
for one particular run of the initial data of~\eqref{phiQz}.

\subsection{The Nature of the Instability}
\label{sec:instability}

One can ask whether this instability results from the particular nature
of AdS spacetime or instead just the fact that one is evolving in an
essentially bounded domain. Beyond the perturbative analysis, one
can imagine sending a scalar pulse down an elevator shaft equipped
with a mirror. It will
blueshift as it gets deeper, but upon return to its original height,
it would obtain its original form only to first order. At higher order,
one considers not just the gravity of the Earth, but also the self-gravity
of the pulse which will act to compress the pulse to its center.

One could confirm this picture by numerically evolving a scalar pulse
in an asymptotically flat space but with a reflecting boundary at some
radius, as done in~\cite{Maliborski:2012gx}. Instead, we do something similar but with our same code. That is,
we enforce reflecting boundary conditions on the scalar field at some
$x_{\rm finite}<\pi/2$
\begin{equation}
\phi_i(x_{\rm finite},t)=0\,,\qquad  \Pi_i(x_{\rm finite},t)=0\,.
\end{equation}
The value is arbitrary, but by truncating the domain, the scalar pulse
does not ``see" that the spacetime in which it lives is AdS. We choose
$x_{\rm finite}= \pi/4$ and find that pulses behave qualitatively in the same
way, i.e. as they bounce, they shift to shorter wavelengths and eventually
form black holes. We subsequently varied the value of $x_{\rm finite}$
and confirmed that this behavior is robust. Interestingly when $\epsilon$
in initial data of the form~\eqref{phiQz} was decreased below $18.39$,
collapse was not observed even after hundreds of bounces.

\subsection{Boundary Information}
\label{sec:boundary}
Given the AdS-CFT correspondence, it is interesting to consider what
the boundary CFT sees from the bulk. One usually uses a ``dictionary''
to extract various quantities. However, here the boundary stress-energy
tensor is time-independent and just equal to the total mass of the
spacetime, $M$. Of particular relevance is the leading asymptotic
value of the scalar field $\phi_i$ as its behavior corresponds to a quantum operator
on the boundary.  We therefore determine the asymptotic behavior of
each scalar field $\phi_i$ with a polynomial, least-squares fit to
\begin{equation}
\phi_i(\rho,t) = \phi_d^{(i)}(t)\ \rho + \phi_{d+2}^{(i)}(t)\ \rho^3.
\label{eq:asymptotics}
\end{equation}
As a check of our extracted value, we employ from 
ten to thirty points and confirm the results of the relevant quantity,
$\phi_d^{(i)}(t)$ are hardly affected.

\begin{figure}
\centerline{\includegraphics[width=4in]{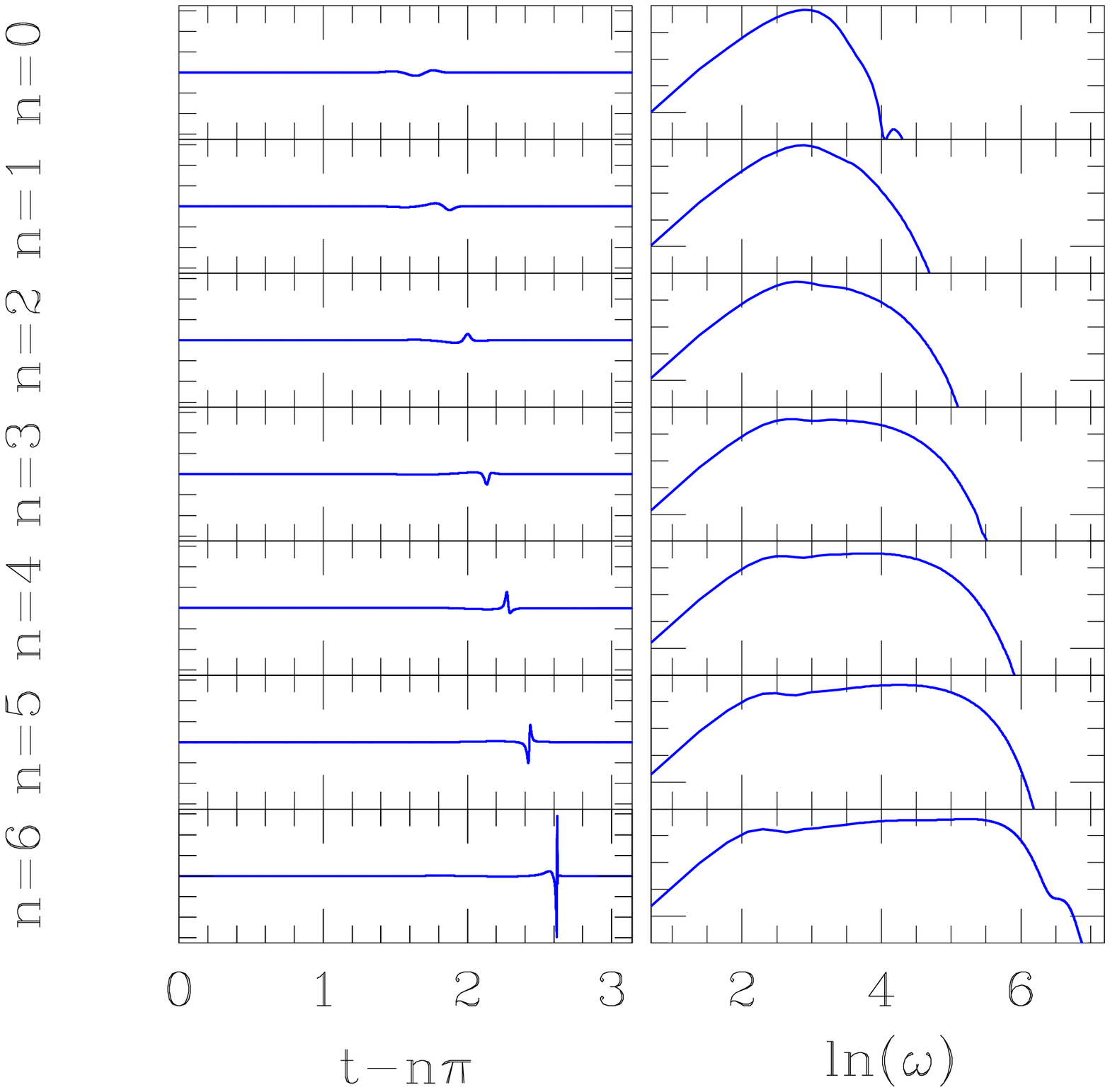}}
\caption{
The behavior of the scalar field at the boundary with successive bounces for the same evolution as in Fig.~\ref{fig:bounces}.
At {\bf left}, the asymptotic value $\phi_3^{(1)}(t)$ from Eq.~(\ref{eq:asymptotics}) is shown for segments of
time of length $\pi$. The pulse sharpens with each ``implosion'' through the origin. On the {\bf right} is shown the Fourier transform of the signal on the left. Shown is the natural logarithm of the amplitude of the Fourier component squared as a function of (the log of)
angular frequency $\omega$. The vertical scale is arbitrary but fixed for all bounces. The transform shows that energy is shifted to higher frequencies.
}
\label{fig:bouncesFFT}
\end{figure}

This asymptotic information is displayed for a particular case
in Fig.~\ref{fig:bouncesFFT}. The left side shows the behavior as a function
of time with each successive bounce along with the Fourier power spectrum.
This spectrum is computed using the FFTW library~\cite{fftw} acting on $\phi_3^{(1)}(t)$ 
once it has been interpolated onto a uniform grid over $t$ (recall the use of AMR
produces nonuniform in time information for the asymptotic behavior). The power is computed
as simply the square of the (complex) Fourier amplitude.


\section{Boundary CFT  perspective of the gravitational collapse}
\label{sec:cft}

The formation of an apparent horizon (AH) in AdS gravitational collapse is holographically dual 
to the thermalization of an initial CFT state corresponding to the initial condition of the 
gravitational evolution. To fully characterize the  gravitational dynamics
from the boundary perspective, one would have to compute an infinite set of local correlation 
functions\footnote{Also, presumably, the nonlocal observables such as Wilson loops.}.
Here, we focus on the simplest local observables, namely $\langle \calo_d^{(i)}\rangle \propto \phi_d^{i}$.
Recall that the one-point function of the boundary stress-energy tensor is constant during the 
evolution (see \eqref{vevs}), and thus does not carry any information about AH formation. 
The same applies to the conserved $U(1)$ charge and so its role is to identify (select) 
dynamical thermalization trajectories.

\subsection{Weakly nonlinear perturbations with global charge in AdS}\label{nonlinear}

Following~\cite{Bizon:2011gg} (see also~\cite{Dias:2011ss}) we consider the solution of \eqref{kg}-\eqref{consteq},
perturbative in the bulk scalar amplitudes $\e$
\begin{equation}
\phi_i=\sum_{j=0}^\infty \e^{2j+1}\ \phi_{i,2j+1}\,,\qquad A=1-\sum_{j=1}^\infty \e^{2j}\ A_{2j}\,,\qquad
\dd=\sum_{j=1}^\infty \e^{2j}\ \dd_{2j}\,,
\eqlabel{pert}
\end{equation}  
where $\phi_{i,2j+1}\,, A_{2j}\,, \dd_{2j}$ are functions of $(t,x)$. It is convenient to decompose 
these functions in terms of a complete basis. A natural basis is provided by 
the $AdS_{d+1}$ massless scalar eigenvalues and eigenfunctions (which we refer to from now 
on as {\it oscillons}) 
\begin{equation}
\w_j=d+2j\,,\ \ e_j(x)=d_j\ \cos^d x\  _2 F_1 \left(-j,d+j,\frac d2, \sin^2 x\right)\,,\ \ 
j=0,1,\cdots\,,
\eqlabel{ossei}
\end{equation}
where $d_j$ are normalization constants such that 
\begin{equation}
\int_0^{\pi/2} dx\ e_i(x)e_j(x) \tan^{d-1} x=\dd_{ij}\,.
\eqlabel{eiej}
\end{equation} 
A remarkable observation of~\cite{Bizon:2011gg} was that initial conditions which represent
at a linearized level (at order $\calo(\e)$ ) a superposition of several 
oscillons with different index $j$
appear to be unstable at time scales $t_{\rm instability}\sim \calo(\e^{-2})$; on the other hand,  nonlinear 
effects of a single oscillon do not lead to destabilization. Specifically, the instabilities occur 
whenever oscillons with indices\footnote{The indices could be repeated.}  $\{j_1,j_2,j_3\}$ 
are present at order $\calo(\e)$, while the oscillon with index $j_r$, such that 
\begin{equation}
\w_{j_r}=\w_{j_1}+\w_{j_2}-\w_{j_3}\,,
\eqlabel{instcond}   
\end{equation}
is not excited at this order.

In what follows we argue that the basic conclusion of~\cite{Bizon:2011gg} remains 
valid even for configurations with $Q\ne 0$. We point out though an interesting twist: 
although any {\it finite} number of distinct-index oscillons excited at order $\calo(\e)$ leads to nonlinear instabilities 
at order $\calo(\e^3)$, $\calo(\e)$ initial configurations with an {\it infinite} number of distinct-index oscillons 
remain (formally) stable at order $\calo(\e^3)$.

While it is straightforward to carry out the analysis in 
full generality, to be maximally explicit we discuss the case of $d=3$ and a superposition 
of two lowest-frequency oscillons at a linearized level, namely $e_0$ and $e_1$. 
We discuss different cases: 
\begin{itemize}
\item a single neutral oscillon 
\item a superposition of two neutral oscillons
\item a single charged oscillon
\item a superposition of two charged oscillons 
\item a superposition of a charged and a neutral oscillon.
\end{itemize} 
Lastly, we comment on nonlinear instabilities at $\calo(\e^3)$ arising from different 
channels \eqref{instcond}, \ie, different combinations of $\{j_1,j_2,j_3\}$ resulting in the 
same $\w_{j_r}$). 

To distinguish neutral and charged initial conditions it is convenient to express the 
charge $Q$ given by \eqref{qinit} as 
\begin{equation}
Q=4\pi^2 \int_0^{\pi/2} dx\ \tan^2 x\ A^{-1}(t,x)\ e^{\dd(t,x)}\ \left(\del_t\phi_1(t,x) 
\phi_2(t,x)-\del_t\phi_2(t,x) 
\phi_1(t,x)\right)\,. 
\eqlabel{qweak}
\end{equation}   

\subsubsection{A single neutral oscillon}\label{c1}

Here we take 
\begin{equation}
\phi_2(t,x)\equiv\phi_1(t,x)\,,\qquad \phi_1(0,x)=\e\ e_0(x)+\calo(\e^3)\,,\qquad \del_t\phi_1(0,x)=0\,.
\eqlabel{bizoninit}
\end{equation}
Clearly, in this case $Q=0$. Requiring normalizability and regularity of $\phi_1$, to order $\calo(\e^3)$ 
(where first nonlinear effects appear), we find
\begin{equation}
\begin{split}
\phi_1=&\e\ \biggl[  e_0(x)\ \cos\left(\left(3-\frac{135}{4\pi}\e^2\right)t\right)\biggr]+
\e^3\ \biggl[F_{3,3}(x)\ \cos(3t)\\
&+F_{3,9}(x)\ \cos(9t)\biggr]+\calo(\e^5) \,,  
\end{split}
\eqlabel{case1}
\end{equation}
with 
\begin{equation}
\begin{split}
F_{3,3}=&\frac{3\sqrt{2}\cos^3 x}{\pi^{3/2}}
\biggl(12\cos^8 x-88\cos^6x+108\cos^4x-63\cos^2x+63\pi^2\\
&-252x^2
-252x\cot x(2-\cos^2x)
\biggr)\,,
\end{split}
\eqlabel{f33}
\end{equation}
\begin{equation}
\begin{split}
F_{3,9}=&\frac{4\sqrt2}{\pi^{3/2}} \cos^9x (9\cos^2x-4)\,.
\end{split}
\eqlabel{f39}
\end{equation}
Notice that in \eqref{case1} we absorbed a term linearly growing in time 
\[
\propto \e^3 t \sin(\w_0 t)
\] 
into $\calo(\e^2)$
shift of the leading-order oscillon frequency $\w_0$:
\begin{equation}
\w_0\to w_0-\frac{135}{4\pi}\e^2\,.
\eqlabel{shift}
\end{equation}
Obviously, we could do so because an oscillon with such a frequency has already been present in 
the initial condition \eqref{bizoninit}. For this initial configuration 
the instability condition \eqref{instcond} is satisfied only for $j_1=j_2=j_3=j_r=0$.

\subsubsection{A superposition of two neutral oscillons}\label{c2}
Consider now a slightly more general (neutral) initial condition
\begin{equation}
\phi_2(t,x)\equiv\phi_1(t,x)\,,\qquad \phi_1(0,x)=\e\ (e_0(x)+e_1(x))+\calo(\e^3)\,,\qquad \del_t\phi_1(0,x)=0\,.
\eqlabel{bizon2init}
\end{equation}
Once again, in this case $Q=0$. Requiring normalizability and regularity of $\phi_1$, to 
order $\calo(\e^3)$, we find
\begin{equation}
\begin{split}
\phi_1=&\e\ \biggl[  e_0(x)\ \cos\left(\left(3-\frac{335}{2\pi}\e^2\right)t\right)
+e_1(x)\ \cos\left(\left(5-\frac{1519}{6\pi}\e^2\right)t\right)
\biggr]+\\
&\e^3\ \biggl[\sum_{k=1}^{8} F_{3,2k-1}(x)\ \cos((2k-1)t)
+\frac{\sqrt{6}\pi}{105}e_2(x)\ t\ \sin(7 t)\biggr]+\calo(\e^5)   \,,
\end{split}
\eqlabel{case2}
\end{equation}
where $F_{3,2j+1}(x)$ are some analytically determined functions\footnote{We  omit their
explicit expressions to keep formulae readable.}.
Here, we have three different terms at order $\calo(\e^3)$, which grow linearly with time
\begin{equation}
\propto \e^3 t\times \biggl\{\cos(\w_0 t)\,,\ \cos(\w_1 t)\,,\ \sin(\w_2 t)\biggr\}\,.
\eqlabel{linear2}
\end{equation}
The presence of $j=\{0,1\}$ oscillons in order $\calo(\e)$ initial conditions
allows us to absorb the first two terms into the shifts of 
the leading-order oscillon frequencies
\begin{equation}
\w_0\to w_0-\frac{335}{3\pi}\e^2\,,\qquad \w_1\to w_1-\frac{1519}{6\pi}\e^2\,.
\eqlabel{shift2}
\end{equation}
We cannot do the same with the remaining term in \eqref{linear2} ---
for this to happen $\phi_1(0,x)$ must contain a term $\propto \e\ e_2(x)$. 
Of course, the presence of $e_2(x)$ at order $\calo(\e)$ in the initial conditions,
while eliminating $\e^3 t\times \sin(\w_2 t)$ term, would generate new resonances 
at $j>2$. Thus, the absence of linearly growing with time terms at order $\calo(\e^3)$
necessitates the excitation of {\it all} oscillons at order $\calo(\e)$ in the
initial condition. In what follows we show that $\calo(\e^3)$ instabilities
can be removed by shifting the leading-order oscillon frequency 
(provided the relevant excitation is present at order $\calo(\e)$), 
even if they arise from different channels \eqref{instcond}.

\subsubsection{A single charged oscillon}\label{c3}
For charged initial data we take
\begin{equation}
\begin{split}
&\phi_1(0,x)=\e e_0(x)+\calo(\e^3)\,,\qquad \del_t\phi_1(0,x)=0\,,\\
&\phi_2(0,x)=0\,,\qquad \del_t\phi_2(0,x)=\e \w_0\ e_0(x)+\calo(\e^3)\,,
\end{split}
\eqlabel{boson1}
\end{equation}
where (see \eqref{qweak})
\begin{equation}
Q=-12\pi ^2\e^2+\calo(\e^4)\,.
\eqlabel{qweak3}
\end{equation}
Initial data \eqref{boson1} is a direct {\it charged} generalization 
of a single-oscillon initial data \eqref{bizoninit}.
In this case, to order $\calo(\e^3)$, we find
\begin{equation}
\begin{split}
&\phi_1=\e \biggl[e_0(x)\cos\left(\left(3-\frac{63}{2\pi}\e^2\right)t\right)\biggr]+\e^3\ \biggl[
F_{3,3}^1(x)\ \cos(3t)\biggr]+\calo(\e^5)\,,\\
&\phi_2=\e \biggl[e_0(x)\sin\left(\left(3-\frac{63}{2\pi}\e^2\right)t\right)\biggr]+\e^3\ \biggl[
F_{3,3}^2(x)\ \sin(3t)\biggr]+\calo(\e^5)\,.
\end{split}
\eqlabel{case3}
\end{equation}
All the linearly with time growing terms are absorbed into the higher-order shift of $\w_0$.

\subsubsection{A superposition of two  charged oscillons}\label{c4}
For charged initial data we take
\begin{equation}
\begin{split}
&\phi_1(0,x)=\e (e_0(x)+e_1(x))+\calo(\e^3)\,,\qquad \del_t\phi_1(0,x)=0\,,\\
&\phi_2(0,x)=0\,,\qquad \del_t\phi_2(0,x)=\e (\w_0\ e_0(x)+\w_1\ e_1(x))+\calo(\e^3)\,,
\end{split}
\eqlabel{boson2}
\end{equation}
where (see \eqref{qweak})
\begin{equation}
Q=-32\pi ^2\e^2+\calo(\e^4)\,.
\eqlabel{qweak4}
\end{equation}
Initial data \eqref{boson2} is a direct {\it charged} generalization 
of the two-oscillon initial data \eqref{bizon2init}.
In this case, to order $\calo(\e^3)$, we find
\begin{equation}
\begin{split}
&\phi_1=\e \biggl[e_0(x)\cos\left(\left(3-\frac{159}{\pi}\e^2\right)t\right)+
e_1(x)\cos\left(\left(5-\frac{2201}{9\pi}\e^2\right)t\right)\biggr]\\
&+\e^3\ \biggl[
\sum_{k=1}^4F_{3,2k-1}^1(x)\ \cos((2k-1)t)+\frac{50\sqrt{6}}{3\pi}e_2(x)\ t\ \sin(7t)\biggr]+\calo(\e^5)\,,\\
&\phi_2=\e \biggl[e_0(x)\sin\left(\left(3-\frac{159}{\pi}\e^2\right)t\right)+
e_1(x)\sin\left(\left(5-\frac{2201}{9\pi}\e^2\right)t\right)\biggr]\\
&+\e^3\ \biggl[
\sum_{k=1}^4F_{3,2k-1}^2(x)\ \sin((2k-1)t)-\frac{50\sqrt{6}}{3\pi}e_2(x)\ t\ \cos(7t)\biggr]+\calo(\e^5)\,.\\
\end{split}
\eqlabel{case4}
\end{equation}
Parallel to the discussion in Section~\ref{c2}, the absence of leading order $\propto \e  e_2(x)$
modes in the initial condition \eqref{boson2} leads to growing resonance terms
\begin{equation}
\propto \e^3 t\times \biggl\{\cos(\w_2 t)\,,\ \sin(\w_2 t)\biggr\}\,.
\eqlabel{linear4}
\end{equation}

\subsubsection{A superposition of a charged oscillon with a neutral one}\label{c5}
For charged initial data we take
\begin{equation}
\begin{split}
&\phi_1(0,x)=\e (e_0(x)+e_1(x))+\calo(\e^3)\,,\qquad \del_t\phi_1(0,x)=0\,,\\
&\phi_2(0,x)=0\,,\qquad \del_t\phi_2(0,x)=\e \w_0\ e_0(x)+\calo(\e^3)\,,
\end{split}
\eqlabel{boson1os1}
\end{equation}
where (see \eqref{qweak})
\begin{equation}
Q=-12\pi ^2\e^2+\calo(\e^4)\,.
\eqlabel{qweak5}
\end{equation}
In this case, to order $\calo(\e^3)$ we find
\begin{equation}
\begin{split}
&\phi_1=\e \biggl[e_0(x)\cos\left(\left(3-\frac{787}{8\pi}\e^2\right)t\right)+
e_1(x)\cos\left(\left(5-\frac{1843}{12\pi}\e^2\right)t\right)\biggr]\\
&+\e^3\ \biggl[
\sum_{k=1,k\ne 5}^8F_{3,2k-1}^1(x)\ \cos((2k-1)t)+\frac{35\sqrt{6}}{4\pi}e_2(x)\ t\ \sin(7t)\biggr]+\calo(\e^5)\,,\\
&\phi_2=\e \biggl[e_0(x)\sin\left(\left(3-\frac{153}{2\pi}\e^2\right)t\right)\biggr]
+\e^3\ \biggl[
\sum_{k=1,k\ne 5}^7F_{3,2k-1}^2(x)\ \sin((2k-1)t)\\
&-\frac{75}{8\pi}e_1(x)\ 
t\ \cos(5t)
+\frac{5\sqrt{6}}{12\pi}e_2(x)\ t\ \cos(7t)\biggr]+\calo(\e^5)\,.
\end{split}
\eqlabel{case5}
\end{equation}
Parallel to the discussion in Section \ref{c2}, the absence of leading order $\propto \e  e_2(x)$
modes in the initial condition \eqref{boson1os1} for both $\phi_1$ and 
$\phi_2$ leads to growing resonance terms
\begin{equation}
\propto \e^3 t\times \biggl\{\cos(\w_2 t)\,,\ \sin(\w_2 t)\biggr\}\,.
\eqlabel{linear5}
\end{equation}
Likewise, 
the absence of a leading order $\propto \e  e_1(x)$
mode in the initial condition~\eqref{boson1os1} for $\phi_2$ 
leads to a growing resonance term
\begin{equation}
\propto \e^3 t\ \cos(\w_1 t)\,.
\eqlabel{linear55}
\end{equation}

\subsubsection{Multi-channel instabilities at $\calo(\e^3)$ and their removal}
Previously, we considered the linear (order $\calo(\e)$) superposition of two  neutral/charged oscillons and 
identified instabilities at order $\calo(\e^3)$ arising from the  resonance term with index $j_r$ 
(see \eqref{instcond}). Furthermore, we showed that if an oscillon $e_{j_r}(x)$ is present in the initial 
data at order $\calo(\e)$, these instabilities can be removed with an appropriate $\calo(\e^2)$  
shift of the oscillon frequency
\begin{equation}
\w_{j_r}\to \w_{j_r}+\e^2 \w_{j_r}^{(1)}\,.
\eqlabel{shiftj}
\end{equation}
Thus, it appears that to avoid $\calo(\e^3)$ instabilities, initial data at order $\calo(\e)$ must contain 
all oscillons. 

In all examples discussed, only a single instability channel was present.  
If all oscillons are excited at order $\calo(\e)$, the same resonance frequency $w_{j_r}$ \eqref{instcond} would arise from 
different channels. We argue here that all such multi-channel instabilities can still be 
removed with a single resonance frequency shift \eqref{shiftj}. To be maximally explicit, consider a neutral, order  
$\calo(\e)$, initial data containing oscillons $j=\{0,1,2,3\}$
\begin{equation}
\phi_2(t,x)\equiv\phi_1(t,x)\,,\qquad \phi_1(0,x)=\e\ \sum_{j=0}^3 A_j e_j(x)+\calo(\e^3)\,,\qquad \del_t\phi_1(0,x)=0\,,
\eqlabel{multi}
\end{equation}
where $A_j$ are generically different amplitudes. Consider a resonance index $j_r=2$. From \eqref{instcond} 
there are 7 distinct instability channels:
\begin{equation}
\begin{split}
&(1):\ \w_{j_r}=\w_{2}+\w_0-\w_0\,,\qquad (2):\  \w_{j_r}=\w_{2}+\w_1-\w_1\,,\\
&(3):\  \w_{j_r}=\w_{2}+\w_2-\w_2\,,\qquad (4):\ \w_{j_r}=\w_{2}+\w_3-\w_3\,,\\
&(5):\ \w_{j_r}=\w_{1}+\w_1-\w_0\,,\qquad (6):\  \w_{j_r}=\w_{3}+\w_0-\w_1\,,\\
&(7):\  \w_{j_r}=\w_{3}+\w_1-\w_2\,.
\end{split}
\eqlabel{minst}
\end{equation}
Each of the channels in \eqref{minst} would contribute a term that grows linearly with time at order 
$\calo(\e^3)$ to $\phi_1$
\begin{equation}
\begin{split}
&\dd\phi_1^{\rm resonance}=\e^3 t\sin(\w_2 t)\   e_2(x)\ \frac1\pi\biggl(
\frac{60613}{120} A_7^3+\frac{35\sqrt{6}}{2} A_3 A_5^2 +\frac{189\sqrt{5}}{4} A_3 A_9 A_5 
\\
&+\frac{50607}{40} A_9^2 A_7+\frac{1090}{3} A_5^2 A_7+\frac{1797}{20} A_3^2 A_7
+\frac{159\sqrt{30}}{2} A_5 A_7 A_9 
\biggr)\,.
\end{split}
\eqlabel{combined}
\end{equation}    
A resonance frequency shift \eqref{shiftj} would additionally contribute 
\begin{equation}
\begin{split}
&\dd\phi_1^{\rm shift}=\e^3 t\sin(\w_2 t)\   e_2(x)\ \frac1\pi\biggl( 
A_7 w_{2}^{(1)}
\biggr)\,.
\end{split}
\eqlabel{freshift}
\end{equation}    
Clearly, provided $A_7\ne 0$, all the instabilities \eqref{minst} can be removed, 
\ie 
\[
\dd\phi_1^{\rm resonance}+\dd\phi_1^{\rm shift}=0\,,
\]
for suitably adjusted $w_{2}^{(1)}$. 

Straightforward, but quite tedious, considerations show that  for a generic neutral initial condition 
\begin{equation}
\phi_2(t,x)\equiv\phi_1(t,x)\,,\qquad \phi_1(0,x)=\e\ \sum_{j=0}^\infty A_j e_j(x)+\calo(\e^3)\,,\qquad \del_t\phi_1(0,x)=0\,,
\eqlabel{multiged}
\end{equation}
the presence of  resonance contributions from  channels 
\begin{equation}
\w_{j_r}=\w_{j_1}+\w_{j_2}-\w_{j_3}\,,
\eqlabel{reschan}
\end{equation}
can be eliminated by the following resonance frequency shift 
\begin{equation}
\w_{j_r}\to \w_{j_r}+ \e^2\ \sum_{\{j_1,j_2,j_3\}}\frac{A_{j_1}A_{j_2}A_{j_3}}{A_{j_r}}\ \times\  \calo(1)\,,
\eqlabel{resshift}
\end{equation}
where summation occurs over all possible instability channels (triplets $\{j_1,j_2,j_3\}$ satisfying \eqref{reschan}).
The amplitudes $A_j$ are expected to decay fast enough as $j\to \infty $ for the sum in \eqref{resshift} to be convergent\footnote{In~\cite{Buchelboson} we present an explicit example where this is the case.}
(and also not to have an apparent horizon already present at the initial hypersurface). 

%
%
To reiterate, the reason why renormalization of the basic resonance frequencies $\w_{j_r}$ would
remove all instabilities up to order $\calo(\e^3)$ is because
to this order the necessary frequency shifts due to distinct instability
channels~\eqref{reschan} add-up linearly, as explicitly demonstrated
by example~\eqref{multi}. From the latter example
it is also clear that  the 'strength' of each of such
shifts---$\w_{2}^{(1)}$ in equation~\eqref{freshift}---is
inversely proportional to the amplitude of a resonance
frequency at order $\calo(\e)$, \ie,\ $\propto\frac{1}{A_{j_r}}$. Thus, provided all
resonances are excited at $\calo(\e)$ (so that this
`inverse amplitude strength' is finite), the
instabilities can always be eliminated as (schematically)
indicated by~\eqref{resshift}.


Note that while we  discussed neutral initial  conditions, similar considerations also apply for 
initial conditions carrying global charge. Whether or not weakly-nonlinear instabilities can be removed
(or removed under some conditions) beyond order $\calo(\e^3)$ is an interesting open question. 
We emphasize that our numerical simulations, as well as those reported in ~\cite{Bizon:2011gg,Jalmuzna:2011qw}
are outside the weakly nonlinear regime.

\subsection{Global charge and the thermalization time}\label{themalizationtime}

\begin{figure}[t]
\begin{center}
\psfrag{lm}{{$\ln M$}}
\psfrag{lt}{{$\ln \frac{t_{\rm AH}}{t_{\rm crossing}}$}}
  \includegraphics[width=2.5in]{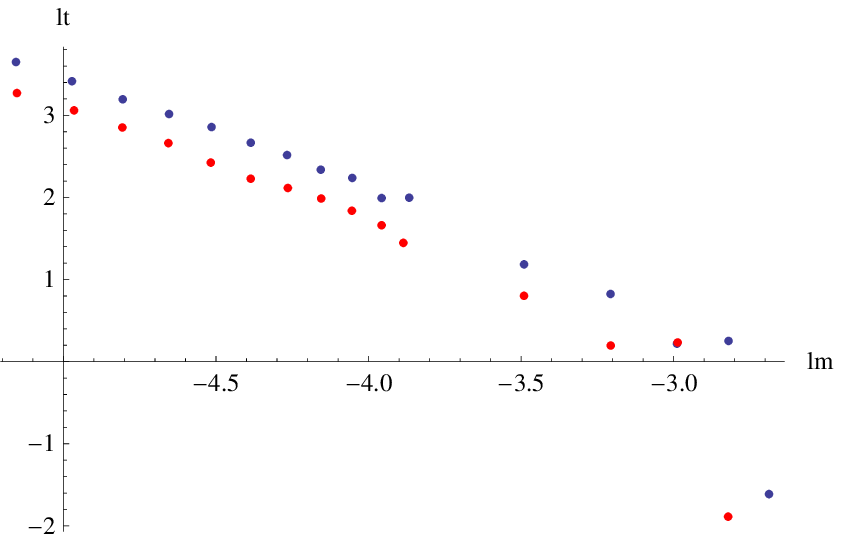}
  ~~~
  \includegraphics[width=2.5in]{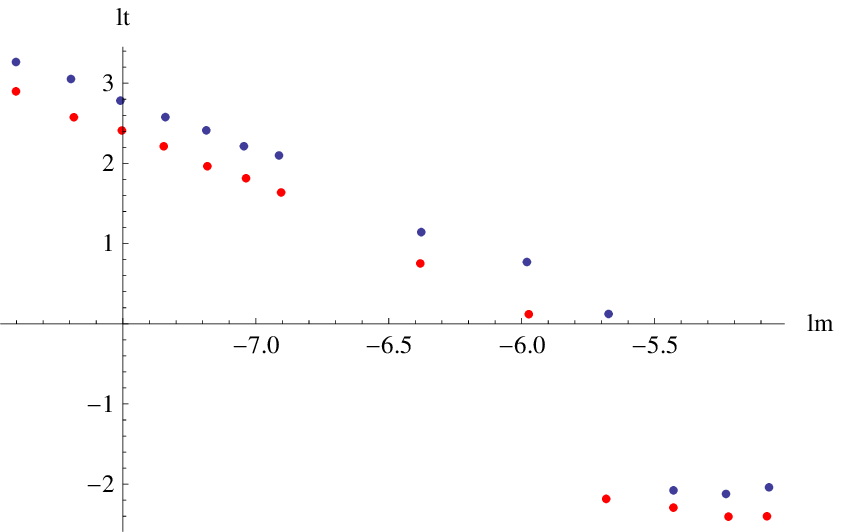}
\end{center}
  \caption{ AH formation time $t_{\rm AH}$ a function of mass $M$
for $d=3$ (left panel) and $d=4$ (right panel) on a log-log plot. 
The blue dots correspond to thermalization of states with $Q=0$;
the red dots correspond to thermalization of states with $Q\ne 0$.  
In both cases, for small $M$ (or $\e$), we find that AH is formed 
earlier than $\propto M^{-1}$ (or $\propto \e^{-2}$) as suggested 
by weakly nonlinear analysis. A global charge $Q$ of the initial configurations 
\eqref{phiQnz} does not extend the thermalization time, compared to 
the same-mass $Q=0$ configurations~\eqref{phiQz}.
}\label{afig5}
\end{figure}

\begin{figure}[t]
\begin{center}
\psfrag{mm}{{$\ln |m_j/m_0|$}}
\psfrag{j}{{$j$}}
  \includegraphics[width=2.5in]{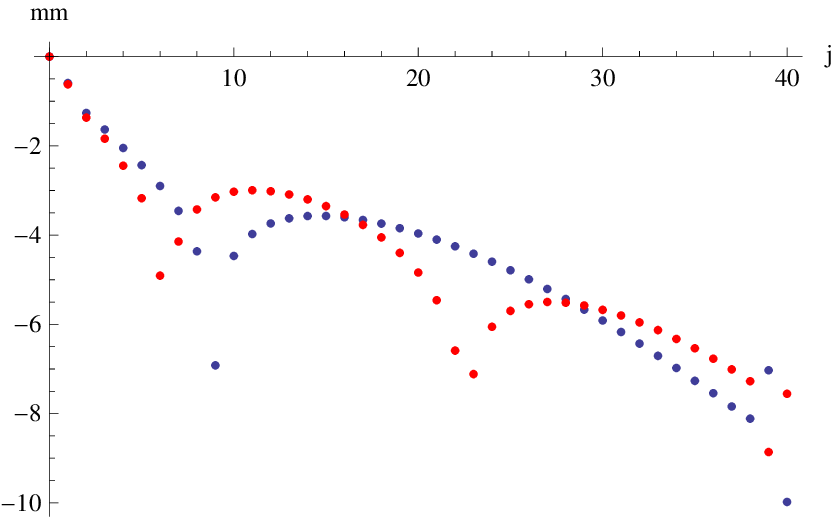}
  ~~~
  \includegraphics[width=2.5in]{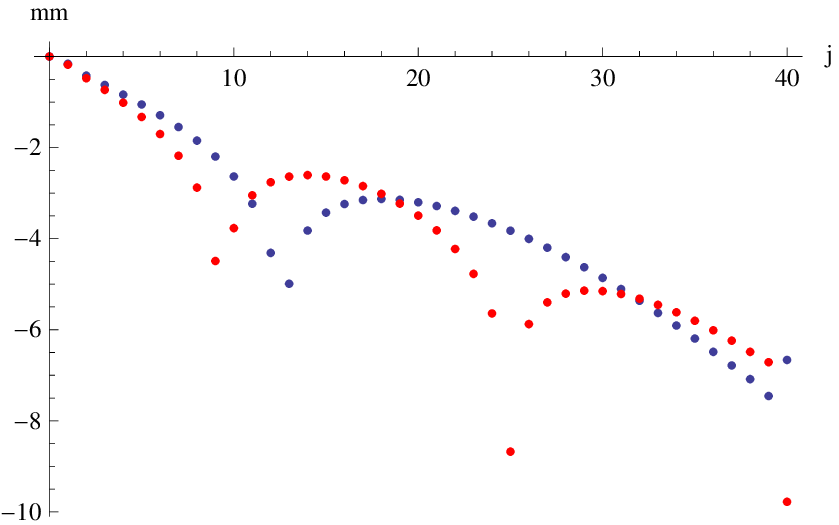}
\end{center}
  \caption{ Decomposition of the mass aspect function $\calm(t=0,x)$ into 
the oscillon basis for two contrasting initial configurations:
    for $\e=10$ neutral~\eqref{phiQz} initial condition (blue dots),
and the charged initial condition~\eqref{phiQnz} with the same mass (red dots).
Left panel: $d=3$, right panel: $d=4$. The different spectra in the oscillon basis 
do not appear to account for the difference in the observed AH formation times. 
}
\label{afig6}
\end{figure}

\begin{figure}[t]
\begin{center}
\psfrag{mm}{{$\ln |m_j/m_0|$}}
\psfrag{tt}{{$\frac{t_{\rm AH}}{t_{\rm crossing}}$}}
\psfrag{e}{{$\e$}}
\psfrag{j}{{$j$}}
  \includegraphics[width=2.5in]{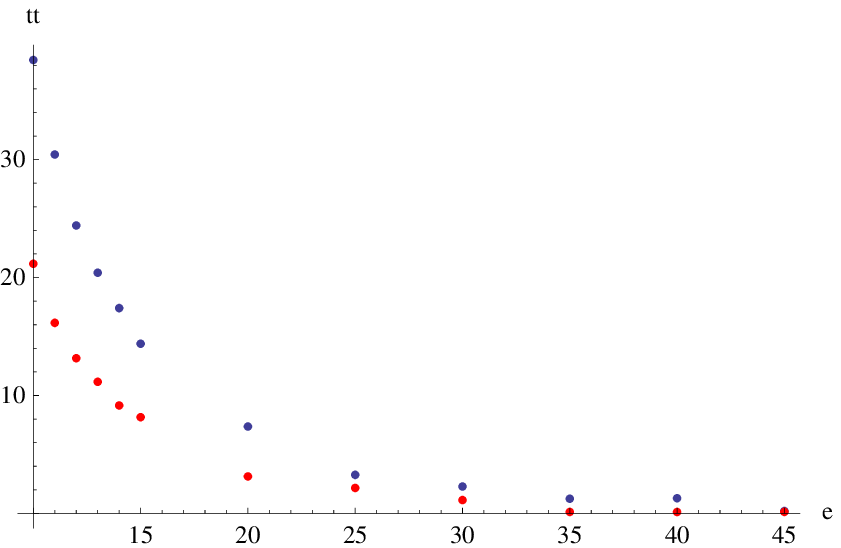}
  ~~~
  \includegraphics[width=2.5in]{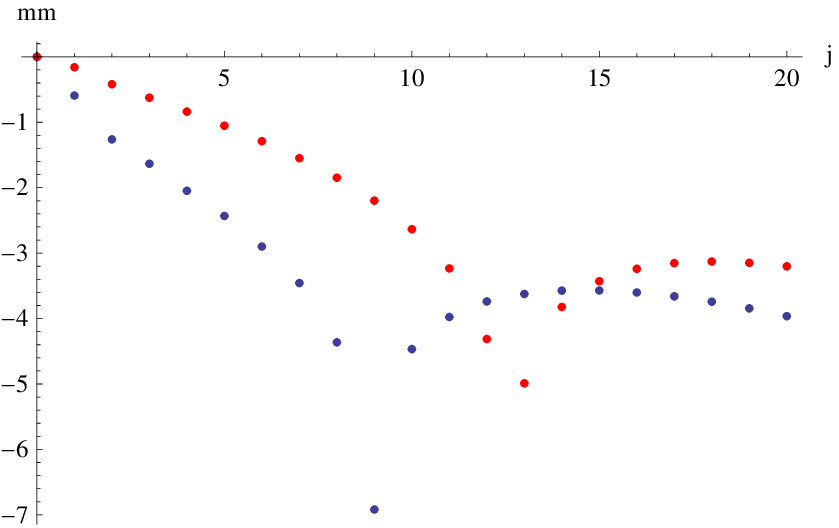}
\end{center}
  \caption{The AH formation time noticeably decreases with increasing 
boundary space-time dimension $d$ (left panel). 
Blue dots represent $d=3$ and red dots $d=4$. This decrease can be 
explained by the broadening of the oscillon decomposition 
 spectrum (see \eqref{osccmass}) of the mass aspect function
corresponding to initial conditions for different $d$ (right panel).
}\label{afig7}
\end{figure}

In global $AdS_d$, the time it takes for a null geodesic
to cross
the space-time is 
\begin{equation}
t_{\rm crossing}=\pi\,.
\eqlabel{tcross}
\end{equation}  
In Fig.~\ref{afig5} we present the ratio of the AH formation time 
$t_{\rm AH}$ to $t_{\rm crossing}$ as a function of mass $M$. The blue dots 
represent collapse from initial condition~\eqref{phiQz}, and 
the red dots correspond to collapse from initial condition~\eqref{phiQnz}.
We find that an AH is forming 
earlier for $Q\ne 0$ initial conditions with the same mass
as the corresponding $Q=0$ initial conditions. 
This difference in collapse times could potentially be attributable to either: (i) that one has a global
charge while the other is neutral, or (ii) that the spectral content of the two initial conditions are
different.
To understand which of these is responsible, 
we decompose the mass aspect function \eqref{massfunction} into the oscillon basis \eqref{ossei}
\begin{equation}
\calm(t=0,x)=\sum_{j=0}^\infty\ m_j\ e_j(x) \,.
\eqlabel{osccmass}
\end{equation}   
The spectral content for the two initial conditions (both with the same mass)
is presented in Fig.~\ref{afig6}. The spectrum of the charged configuration 
does {\em not} show a
noticeably broader bandwidth or enhanced amplitudes 
at high frequencies when compared to the spectrum of the neutral configuration. Thus, it 
appears that the acceleration of the 
collapse is a dynamical effect related to the class of initial conditions~\eqref{phiQnz}.

In a similar fashion, we find significant differences in collapse times when changing the spacetime dimensions $d$.
However, in this case the spectral content of the initial mass aspect function 
{\em does} account for the difference in AH formation time.
The left panel in Fig.~\ref{afig7} compares AH formation 
times for $d=3$ (blue dots) and $d=4$ (red dots). The spectral content of the 
corresponding mass aspect function is shown in the right panel. Because higher dimensional 
oscillons are more localized in the $AdS_{d+1}$ center with increasing $d$, there is 
larger overlap 
of the initial profile \eqref{phiQz} with 
higher modes for larger $d$, resulting in broadening of the spectral bandwidth.

\subsection{Bulk vs boundary turbulence}\label{turbulence}

A motivating question behind this study wonders what the CFT sees
of this instability on the boundary. Because the only observables
on the boundary we consider are the (unchanging) total mass $M$ and the asymptotic
behavior of the scalar fields $\phi^{(i)}_d(t)$ (see~\eqref{uv})
(or equivalently $\langle\calo_d^{(i)}\rangle$, see \eqref{vevs}),
we examine the spectral content of the latter.

Fig.~\ref{fig:bouncesFFT} displays snapshots of the scalar field for successive bounces
during an evolution. The initial pulse becomes increasingly more compact,
approaching something like a Kronecker delta function. As it does so,
the FFT of each pulse similarly approaches the FFT of a delta function,
in particular, a flat, plateau-like spectrum. In general, if we
define the effective bandwidth of the power spectrum as the frequency
at which the plateau ceases, then we observe this bandwidth to increase
rapidly during an evolution. The only exception appears, for some runs,
during the last few bounces, during which the bandwidth decreases modestly
from its maximum. However, we speculate that this behavior can be
attributed to the nonlinearities of gravity near black hole formation.

Before discussing the behavior of the bandwidth, we should comment
on the falloff behavior of the power spectrum. It is customary in
studies of turbulence in fluid dynamics to observe the rate of falloff
(the slope on a log-log plot). In particular, a famous result of
Kolmogorov expects a slope in the {\em energy spectra} of $-5/3$ within the inertial
regime of  turbulent flows (independent of spatial dimension $d$).
Here, one might consider the scalar field, in analogy with a fluid, as
a ``gas'' of oscillons in which structure transitions to increasingly higher 
 frequencies. An analysis of the energy spectra over the
oscillon basis in the bulk in~\cite{Maliborski:2012gx} finds a fall-off
at the rate of $-1.22$. As well, an analysis of the Fourier spectrum over
the time series of the Ricci scalar at the origin in~\cite{deOliveira:2012dt}
reports a slope of $-1.61$, seemingly close to that of Kolmogorov. However,
the Ricci scalar is an energy density and thus the expected scaling is not $-5/3$. Indeed, we can
retrace Kolmogorov's argument for such quantity by the following dimensional considerations.
The Fourier transform of $\Pi^2$ (which scales as ${\cal E} \propto T^{-2}$) scales as 
$\hat {\cal E} \propto L T^{-2}$ for
a typical wavelength $L$. If turbulent behavior transfers energy to different scales,
for $\Pi^2$  it does
so with an efficiency 
 $\eta$ which has dimensions of $T^{-3}$ (transfer per unit time).
Thus, one would expect these quantities to be related by
\begin{equation}
\hat {\cal E} \equiv \left [ \frac{L}{T^2} \right ] = \eta^p \omega^{\alpha} \equiv 
\left [ \frac{1}{T^3} \right ]^p \left[ \frac{1}{L} \right]^{-\alpha}\,,
\label{kolmogorov}
\end{equation}
where $p$ and $\alpha$ are (so far) arbitrary powers. However, for~\eqref{kolmogorov} to
be dimensionally consistent, then we must have $p=2/3$ and
and $\alpha = -1$.

In evaluating what our data says about the falloff, we have to know which
parts of the spectrum represent true physics as opposed to numerical noise.
Straightforward unigrid convergence tests suggests that we can roughly
trust the spectra up to a bit better than $\omega \approx 500$ (i.e. the 
Nyquist frequency $\omega_N = \pi/(2 \Delta x)$ with $\Delta x$ the grid spacing). 
However,
in general, we run with AMR and can resolve locally enormously better than
any of our unigrid evolutions. One issue is that while AMR is very
efficient locally, its performance on global measures is  difficult
to evaluate.  Instead, we carryout some runs with different AMR parameters
and evaluate where the spectrum changes significantly. 

Additionally, AMR introduces its own
noise into the asymptotic behavior of the scalar field, especially for
higher dimensions ($d\ge 6$). This noise appears as very small kinks
in the time series data which contribute noise that behaves as $\propto \omega^{-2}$ 
(such that the power fall-off from this noise source is $-4$).

We also note two other sources of noise in the spectrum. In order to
carryout a Fourier decomposition, we first interpolate to a grid uniform
in $t$. Comparing FFTs with either linear or cubic interpolation, we find
this noise is generally small, only significant at quite high frequencies and not
strongly dependent on the interpolation order.
The final source of noise arises from potential mismatches in the time-series
data between $\phi^{(i)}_d(n\pi)$  and $\phi^{(i)}_d(\left[n+1\right]\pi)$.
Because the FFT expects a periodic signal, this mismatch appears as a
(quite small) discontinuity producing noise that falls off 
as $\propto \omega^{-1}$ (with a power that falls off with exponent $-2$).

As a practical matter, we find that for different runs, different noise
sources dominate the high-frequency behavior of the power spectrum.
Indeed the only robust features of the various power spectra we have computed
at the AdS boundary
have been the delta-function plateau and high-frequency noise the power of
which falls off either with slope $-2$ or $-4$. In particular, 
to make a convincing argument that some particular fall-off is characteristic
of this weakly nonlinear instability, one needs to demonstrate not only
a feature of the spectrum independent of any noise, but also a feature not
present in the initial spectrum. That is, any characteristic feature
must {\em develop} from the evolution.

\begin{figure}
\centerline{\includegraphics[width=4in]{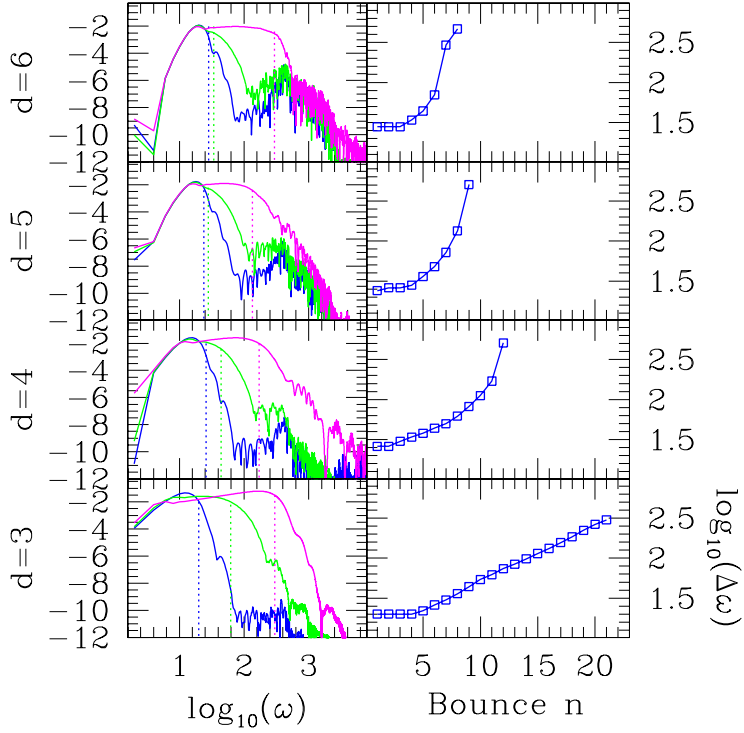}}
\caption{Demonstration of the bandwidth growth with each bounce.
Shown are the results from the evolution of initial data of form~\eqref{phiQz}
with $\epsilon=8$,
$\sigma=0.1$ for dimensions $d=3,4,5$ and $6$.
On the {\bf left} is shown the spectrum of
the first (blue), middle (green), and last (magenta) bounces.
The pulses increasingly demonstrate a plateau-like region indicative
of becoming more like a delta function. At frequencies above
the plateau, unphysical noise is apparent.
On the {\bf right} is shown the bandwidth associated with all pulses.
The bandwidth is obtained as the location at which the spectrum
drops from its maximum by twenty percent.
As noted in the text (see Fig.~\ref{afig7}), the same initial profile for the scalar field
contains higher oscillons in its decomposition with higher
dimensions, 
and therefore one observes the bandwidth of the pulses grow more
quickly than the $d=3$ case.
}
\label{fig:bandwidth}
\end{figure}

At the boundary, the only such feature we find is that of the plateau and we study 
its defining feature, its bandwidth.
In Fig.~\ref{fig:bandwidth}, we display the results of measuring the bandwidth
for one set of evolutions. We show the power spectrum of the first, middle, and
last bounces. From each bounce, we calculate a maximum frequency as the location
where the spectral power falls some threshold below the maximum (typically $20\%$ below the maximum). 
This frequency then serves as an estimate of the bandwidth
of the pulse and is shown on the right side of the figure. As the pulse 
sharpens in time, its spectrum reaches into higher frequencies and the rate
at which it does so is roughly exponential.

\begin{figure}
\centerline{\includegraphics[width=4in]{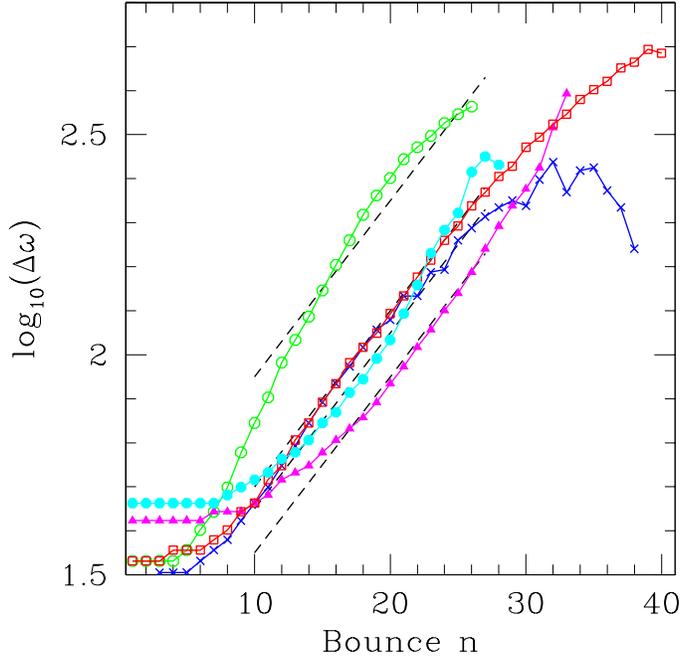}}
\caption{Bandwidth as a function of bounce for a few different evolutions. Choosing evolutions
with roughly 30 bounces with different dimension $d$ and different families, we observe similar,
exponential growth of the bandwidth for some intermediate regime (roughly from bounce 10 to
bounce 30).
Shown are the results from:
(blue crosses)
          initial configuration~\eqref{phiQz} with $\epsilon=10$ in $d=3$,
(green open circles)
     initial configuration~\eqref{phiQnz} with $\epsilon=10$ in $d=3$,
(cyan open circles)
     initial configuration~\eqref{phiQz} with $\epsilon=7$ in $d=6$,
(magenta triangles)
     initial configuration~\eqref{phiQz} with $\epsilon=7$ in $d=5$,
 and
(red squares)
          a Gaussian initial profile in $d=3$.
four (black dashed) lines all with the slope $0.04$ but with various $y$-intercepts chosen to align
with the bandwidth data are shown.
}
\label{fig:growthrate}
\end{figure}

Looking at the right side of  Fig.~\ref{fig:bandwidth}, one notices that the growthrate of
the bandwidth, after an initial quiescent period,  appears to approach a line,
representing exponential growth of the bandwidth. We are thus led to examine the bandwidth
of runs with a relatively large number of bounces, and indeed we find exponential growth
of the bandwidth for these runs, shown in Fig.~\ref{fig:growthrate}.

What is particularly interesting about Fig.~\ref{fig:growthrate} is that these
runs demonstrate what appears to be a common growth rate with slope $0.04$. This
commonality in slope appears despite the fact that these families span different dimensions $d$
and different forms of initial data. What they have in common is that they all form black holes
after between 26 and 39 bounces. Note that the growth rate for these evolutions differs from
that shown for $d=3$ in Fig.~\ref{fig:bandwidth} (with a slope of roughly $0.07$)
which collapses after just 21 bounces.

Another interesting aspect of these bandwidth-vs-bounce plots is that nearly all evolutions
studied terminate at roughly $\log_{10}\left(\Delta \omega\right) \approx 2.7$. Presumably
this reflects that black hole formation occurs for such high-frequencies. We speculate
based on these two features, that a bandwidth plot such as discussed here can be considered as
something of a phase-space picture for this system. At the top of the plot, is black
hole formation. Low on the left side, evolutions not yet consisting of a black hole evolve
through some quiescent period. Depending on how weak the initial data, they enter an
exponentially growing bandwidth regime until they are brought to black hole formation.
Perhaps, the growth rate in this linear regime is set by how low they begin, or, equivalently,
how many bounces they require to achieve the black hole scale. We stress that this
is speculation and will require more runs to test whether indeed common growth rates
occur for disparate evolutions that share the same number of bounces. Similarly,
this phase space picture may require some other basis, such as the oscillon basis, over
which to compute the bandwidth.

\begin{figure}
\begin{center}
\epsfig{file=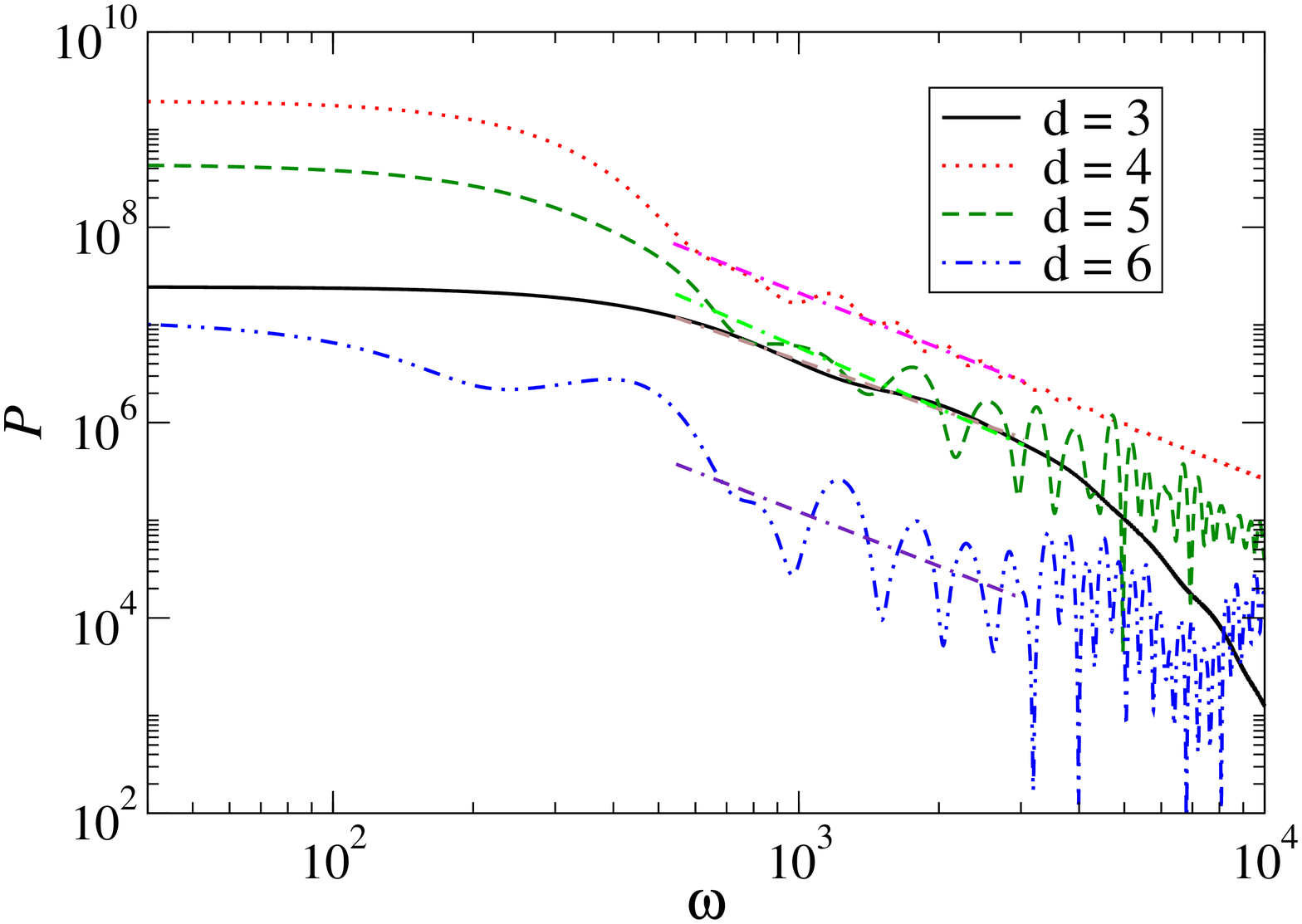,height=10.4cm,angle=-90}
\caption{Power spectra of $\Pi^2$ at the origin versus frequency $\omega$ for $d=3,4, 5,6$ as measured
in the last bounce before black hole formation. At frequencies below
$\approx 800$, the apparent plateau behavior is indicative of the sharpened profile
of the pulse. At higher frequencies up to $\omega \approx 3000$, the spectral power
decreases with a slope in the range $\in (-1.8, -2.2)$. 
}
\label{fig:pispectra_last}
\end{center}
\end{figure}

Last, we also monitor the behavior of $\Pi^2$ {\em at the origin} 
which is related to
the energy density (and scales in the same way).
Numerically, the behavior of the scalar fields at the origin produces less noise and hence cleaner power spectra.
Figure~\ref{fig:pispectra_last} presents results for
$d=3, 4, 5,$ and $6$ for the last bounce before black hole formation. At frequencies below 
 $\omega \simeq 800$ a plateau behavior results from the pulse profile being 
quite narrow and approaching a delta-like behavior. Within the range $\omega \in (800,3000)$ 
the slopes measured fall in the interval $\approx (-1.8, -2.2)$, consistent with our previous
Kolmogorov-type argument for the power spectrum of a quantity scaling as energy density (recall that
the {\em power} spectrum is the square of the complex Fourier amplitude and hence should fall-off as $-2\alpha$).

The relatively wide range of this interval is perhaps not surprising given that we lack sharp knowledge
of where the inertial regime lies in the spectrum\footnote{To more easily identify this regime,
one could introduce boundary deformations to continuously pump energy from the boundary into the system
as done in e.g.~\cite{Buchel:2012gw}}. 
On one hand, because energy is only ``injected'' initially, as it shifts to higher frequencies
it pushes the initial lower bound for the inertial frequency higher. On the other hand, at high
frequencies one has a ``dissipative'' scale
determined by the black hole size. However, because energy is also ``spent'' in curving the spacetime, the location
of this dissipative bound is also difficult to determine. Nevertheless, inspection of slope breaks in the low and high frequency
regimes leave an intermediate domain in which the power spectra appears to fall-off with a slope in the range mentioned.
Pushing the analogy with turbulent fluid dynamics even further,
one could speculate that deviations from a slope of $-2$ are not surprising, but instead that they would arise 
much as {\em intermittency} arises in turbulent fluids.
With fluids, intermittency 
introduces deviations from the $-5/3$ slope due to the presence and dynamics of vortices. Here
perhaps
the sharpening pulse as it periodically propagates through the origin could leave a 
similar imprint on the spectrum.


\section{Conclusion}
\label{sec:conclude}

We have reproduced the work of~\cite{Bizon:2011gg,Jalmuzna:2011qw}
and provided details of our numerics and tests. We find that this
instability towards higher frequencies occurs even in a finite-sized
domain with reflecting boundary conditions that would not be expected to
know that the spacetime is asymptotically AdS. This result, along with the similar 
case in asymptotically Minkowski in the work of~\cite{Maliborski:2012gx}, suggests that
the instability is a property not of AdS itself, but instead of gravity
in a bounded domain.

We generalized the work of ~\cite{Bizon:2011gg,Jalmuzna:2011qw} to complex scalar
 collapse in AdS to address 
the question of whether AH formation in global AdS from generic initial conditions 
from the bulk perspective
(or  thermalization trajectories from the boundary CFT perspective ) is sensitive to the presence of a 
conserved global charge. For the class of initial conditions we study (\eqref{phiQz} and \eqref{phiQnz}) 
the answer appears to be ``no.'' Rather, the collapse is affected by the spectral decomposition 
of the initial 
data in the oscillon basis (see \eqref{ossei}). We would like to stress that global charge can 
(and does \cite{Buchelboson}) strongly influence AH formation trajectories in the vicinity 
of stable stationary configurations in AdS. 

We repeated the weakly nonlinear analysis of the gravitational collapse in the oscillon basis,
originally presented in \cite{Bizon:2011gg}. We argued that to the leading order in nonlinearities,
all the instabilities can be removed, provided {\it all} oscillons are excited at the linearized level. 
This conclusion is robust and independent of the global charge. We further argued that, while 
the picture of gravitational collapse as an ideal gas of oscillons is valid in the linearized 
regime, this interpretation is in conflict with the  late-time evolution.

Finally, it is clear that gravitational collapse in AdS (or confined gravity) 
is driven by some focusing mechanism by which the bulk energy density cascades to shorter scales.
We have analyzed this behavior from both bulk and boundary perspectives. In the latter, we have
uncovered an exponential behavior in the growth of the spectral bandwidth of the scalar field. This observation
is also confirmed by examining the behavior at the origin and, interestingly, $\Pi^2$
displays a behavior consistent with a Kolmogorov-like mechanism. However, further investigations
are required to put this observation on firmer footing.

%
%
~\\
\noindent{\bf{\em Acknowledgments:}}
It is a pleasure to thank Oscar Dias, Chad Hanna, Gary Horowitz, Pavel Kovtun, Robert Myers and Jorge Santos
for interesting and helpful discussions.
This work was supported by the NSF (PHY-0969827 to Long Island University)
and NSERC through Discovery Grants (to AB and LL) as well as CIFAR (to LL). 
LL and SLL thank the KITP 
for hospitality where parts of this work were completed.
Research at Perimeter
Institute is supported through Industry Canada and by the Province of Ontario
through the Ministry of Research \& Innovation.
Computations were performed thanks to allocations at the Extreme Science and Engineering 
Discovery Environment (XSEDE),
which is supported by National Science Foundation grant number OCI-1053575 as well as SHARCNET.

\bibliographystyle{utphys}     
\bibliography{./scalar}

\end{document}

%% file: alex2.tex
\ifnum\draftcontrol=1
\tolerance=1000
\fi

\renewcommand\baselinestretch{1.25}
\setlength{\paperheight}{11in}
\setlength{\paperwidth}{8.5in}
\setlength{\textwidth}{\paperwidth-2.4in}     \hoffset= -.3in   
\setlength{\textheight}{\paperheight-2.4in}   \topmargin= -.6in 

\renewcommand\section{\@startsection {section}{1}{\z@}%
                                   {-3.5ex \@plus -1ex \@minus -.2ex}%
                                   {2.3ex \@plus.2ex}%
                                   {\normalfont\large\bfseries}}
\renewcommand\subsection{\@startsection{subsection}{2}{\z@}%
                                   {-3.25ex\@plus -1ex \@minus -.2ex}%
                                   {1.5ex \@plus .2ex}%
                                   {\normalfont\normalsize\bfseries}}
\renewcommand\subsubsection{\@startsection{subsubsection}{3}{\z@}%
                                   {-3.25ex\@plus -1ex \@minus -.2ex}%
                                   {1.5ex \@plus .2ex}%
                                   {\normalfont\normalsize\it}}
\renewcommand\paragraph{\@startsection{paragraph}{4}{\z@}%
                                   {-3.25ex\@plus -1ex \@minus -.2ex}%
                                   {1.5ex \@plus .2ex}%
                                   {\normalfont\normalsize\bf}}

\numberwithin{equation}{section}



\def\ie{{\it i.e.}}
\def\eg{{\it e.g.}}

\def\revise#1       {\raisebox{-0em}{\rule{3pt}{1em}}%
                     \marginpar{\raisebox{.5em}{\vrule width3pt\
                     \vrule width0pt height 0pt depth0.5em
                     \hbox to 0cm{\hspace{0cm}{%
                     \parbox[t]{4em}{\raggedright\footnotesize{#1}}}\hss}}}}

\newcommand\fnxt[1] {\raisebox{.12em}{\rule{.35em}{.35em}}\mbox{\hspace{0.6em}}#1}
\newcommand\nxt[1]  {\\\fnxt#1}

\def\cala         {{\cal A}}
\def\calA         {{\mathfrak A}}
\def\calAbar      {{\underline \calA}}
\def\calb         {{\cal B}}
\def\calc         {{\cal C}}
\def\cald         {{\cal D}}
\def\cale         {{\cal E}}
\def\calf         {{\cal F}}
\def\calg         {{\cal G}}
\def\calG         {{\mathfrak G}}
\def\calh         {{\cal H}}
\def\cali         {{\cal I}}
\def\calj         {{\cal J}}
\def\calk         {{\cal K}}
\def\call         {{\cal L}}
\def\calm         {{\cal M}}
\def\caln         {{\cal N}}
\def\calo         {{\cal O}}
\def\calp         {{\cal P}}
\def\calq         {{\cal Q}}
\def\calr         {{\cal R}}
\def\cals         {{\cal S}}
\def\calt         {{\cal T}}
\def\calu         {{\cal U}}
\def\calv         {{\cal V}}
\def\calw         {{\cal W}}
\def\calz         {{\cal Z}}

\def\complex      {{\mathbb C}}
\def\naturals     {{\mathbb N}}
\def\projective   {{\mathbb P}}
\def\rationals    {{\mathbb Q}}
\def\reals        {{\mathbb R}}
\def\zet          {{\mathbb Z}}

\def\del          {\partial}
\def\delbar       {\bar\partial}
\def\ee           {{\rm e}}
\def\ii           {{\rm i}}
\def\chain        {{\circ}}
\def\tr           {\mathop{\rm Tr}}
\def\Re           {{\rm Re\hskip0.1em}}
\def\Im           {{\rm Im\hskip0.1em}}
\def\id           {{\it id}}

\def\de#1#2{{\rm d}^{#1}\!#2\,}
\def\De#1{{\cald}#1\,}

\def\half{{\frac12}}
\newcommand\topa[2]{\genfrac{}{}{0pt}{2}{\scriptstyle #1}{\scriptstyle #2}}
\def\undertilde#1{{\vphantom#1\smash{\underset{\widetilde{\hphantom{\displaystyle#1}}}{#1}}}}
\def\prodprime{\mathop{{\prod}'}}
\def\gsq#1#2{%
    {\scriptstyle #1}\square\limits_{\scriptstyle #2}{\,}} 
\def\sqr#1#2{{\vcenter{\vbox{\hrule height.#2pt
 \hbox{\vrule width.#2pt height#1pt \kern#1pt
 \vrule width.#2pt}\hrule height.#2pt}}}}
\def\square{%
  \mathop{\mathchoice{\sqr{12}{15}}{\sqr{9}{12}}{\sqr{6.3}{9}}{\sqr{4.5}{9}}}}

\newcommand{\fft}[2]{{\frac{#1}{#2}}}
\newcommand{\ft}[2]{{\textstyle{\frac{#1}{#2}}}}
\def\jsquare{\mathop{\mathchoice{\sqr{8}{32}}{\sqr{8}{32}}
{\sqr{6.3}{9}}{\sqr{4.5}{9}}}}

\newcommand{\wn}{\mathfrak{w}}


\def\a{\alpha}
\def\b{\beta}
\def\l{\lambda}
\def\w{\omega}
\def\dd{\delta}
\def\r{\rho}
\def\c{\chi}
\newcommand{\qq}{\mathfrak{q}}
\newcommand{\ww}{\mathfrak{w}}
\def\p{\phi}
\def\k{\kappa}
\def\hr{\hat{r}}
\def\om{\Omega}
\def\e{\epsilon}
\def\bm{\bar{\mu}}
\def\tc{\tilde{\calc}}
\def\tz{\tilde{z}}
\def\hz{\hat{z}}
\def\t{\tau}

\catcode`\@=12